\documentclass{pasj00}

\color{black} 

\begin{document}
\SetRunningHead{Y.\ Terada {\it et al.}}
{Timing calibration of the HXD on board Suzaku}
\Received{2007/06/16}
\Accepted{2007/08/29}

\title{In Orbit Timing Calibration of the Hard X-Ray Detector \\
on Board Suzaku}

\author{
Yukikatsu \textsc{Terada},\altaffilmark{1},
Teruaki \textsc{Enoto},\altaffilmark{2},
Ryouhei \textsc{Miyawaki},\altaffilmark{2},
Yoshitaka \textsc{Ishisaki},\altaffilmark{3},
}
\author{
Tadayasu \textsc{Dotani},\altaffilmark{4},
Ken \textsc{Ebisawa},\altaffilmark{4},
Masanobu \textsc{Ozaki},\altaffilmark{4},
Yoshihiro \textsc{Ueda},\altaffilmark{5},
Lucien \textsc{Kuiper},\altaffilmark{6},}
\author{
Manabu \textsc{Endo},\altaffilmark{7},
Yasushi \textsc{Fukazawa},\altaffilmark{8},
Tsuneyoshi \textsc{Kamae},\altaffilmark{9},
Madoka \textsc{Kawaharada},\altaffilmark{10},}
\author{
Motohide \textsc{Kokubun},\altaffilmark{4},
Yoshikatsu \textsc{Kuroda},\altaffilmark{7},
Kazuo \textsc{Makishima},\altaffilmark{2,10},
Kazunori \textsc{Masukawa},\altaffilmark{7},}
\author{
Tsunefumi \textsc{Mizuno},\altaffilmark{8},
Toshio \textsc{Murakami},\altaffilmark{11},
Kazuhiro \textsc{Nakazawa},\altaffilmark{2},
Atsushi \textsc{Nakajima},\altaffilmark{7},}
\author{
Masaharu \textsc{Nomach},\altaffilmark{12},
Naoki \textsc{Shibayama},\altaffilmark{7},
Tadayuki \textsc{Takahashi},\altaffilmark{4},
Hiromitsu \textsc{Takahashi},\altaffilmark{8},}
\author{
Makoto S. \textsc{Tashiro},\altaffilmark{1},
Toru \textsc{Tamagawa},\altaffilmark{10},
Shin \textsc{Watanabe},\altaffilmark{4},
Makio \textsc{Yamaguchi},\altaffilmark{7},}
\author{
Kazutaka \textsc{Yamaoka},\altaffilmark{13},
\and, Daisuke \textsc{Yonetoku},\altaffilmark{11}}

\altaffiltext{1}{Department of Physics, Science, Saitama University,
Saitama 338-8570, Japan}
\altaffiltext{2}{Department of Physics, Science, University of Tokyo, 
Tokyo 113-0033, Japan}
\altaffiltext{3}{Department of Physics, Tokyo Metroporitan University,
1-1 Minami-Osawa, Hachioji-si, \\ Tokyo, 192-0397, Japan}
\altaffiltext{4}{Institute of Space and Astronautical Science,
Japan Aerospace Exploration Agency (ISAS/JAXA), \\ Kanagawa 229-8510, Japan}
\altaffiltext{5}{Department of Physics, Kyoto University, 
Oiwake, Shirakawa, Sakyou-ku, \\ Kyoto, 606-8502, Japan}
\altaffiltext{6}{SRON, National Institute for Space Research, 
Sorbonnelaan 2, 3584 CA Utrecht, Netherlands}
\altaffiltext{7}{Mitsubishi Heavy Industries, LTD, 
Higashi-tanaka 1200, Komaki, Aichi, 485-8561, Japan}
\altaffiltext{8}{Department of Physical Science, Hiroshima University,
Hiroshima 739-8526, Japan}
\altaffiltext{9}{Stanford Linear Accelerator Center (SLAC),
2575 Sand Hill Road, Menlo Park, CA 94025, USA}
\altaffiltext{10}{Makishima Cosmic Radiation Laboratory, RIKEN,
2-1, Hirosawa, Wako-shi, \\ Saitama 351-0198, Japan}
\altaffiltext{11}{Department of Physics, Science, Kanazawa University,
Kanazawa Ishikawa 920-1192, Japan}
\altaffiltext{12}{Department of Physics, Science, Osaka University,
Osaka 560-0043, Japan}
\altaffiltext{13}{Department of Physics and Mathematics,
Aoyama Gakuin University, Kanagawa 229-8558, Japan}
\email{terada@riken.jp}


\maketitle

\begin{abstract}
The hard X-ray detector (HXD) on board the X-ray satellite Suzaku
is designed to have a good timing capability with a 61 $\mu$s
time resolution.
In addition to detailed descriptions of the HXD timing system,
results of in-orbit timing calibration and performance of the HXD 
are summarized.
The relative accuracy of time measurements of the HXD event was
confirmed to have an accuracy of $1.9\times 10^{-9}$ s s$^{-1}$ per day, 
and the absolute timing was confirmed to be accurate to 360 $\mu$s or better.
The results were achieved mainly through observations of the Crab pulsar,
including simultaneous ones with RXTE, INTEGRAL, and Swift.
\end{abstract}

\section{Introduction}
\label{section:hxd_intro}
The Hard X-ray Detector (HXD; \cite{hxd2007a,hxd2007b})
on board Suzaku \citep{suzaku2007} is a novel cosmic
hard X-ray instrument working  in the 10 -- 600 keV range.
In addition to the very low background level, 
the wide energy band, and the tightly collimated field of view,
yet another important feature of the HXD is its timing capability.
This is because time variability, periodic or aperiodic,
is an important characteristic of compact cosmic X-ray sources,
including neutron stars, white dwarfs,
galactic black holes, and active galactic nuclei.
Considering the wide range of time scales involved in these variations,
we have required the HXD to have a time resolution of 61 $\mu$s,
and a timing stability of the order of $10^{-9}$ typically for a day.
To fulfill these goals, we carefully designed the timing system of the HXD,
as summarized in section \ref{section:hxd_timing},
and in more detail in \citet{hxd2007a} and \citet{hxd2007b}.

The HXD sensor (HXD-S) consists of 16 identical ``Well units''
working as main detectors \citep{IEEE_uchiyama2001},
and 20 surrounding shield counters
made of Bi$_4$Ge$_3$O$_12$ (BGO) scintillators 
\citep{IEEE_yamaoka2006}.
Each Well unit detects incoming X-rays using
2 mm-thick silicon PIN diodes \citep{IEEE_sugiho2001}
and Gd$_2$SiO$_5$:Ce (hereafter GSO) scintillators,
both surrounded by a common active BGO shield.
The HXD signals from the Well Units (called WEL type data) 
are processed first by an analog electronics package (HXD-AE),
and then by a digital electronics (HXD-DE).

\begin{figure}[bht]
\begin{center}
\FigureFile(0.95\columnwidth,){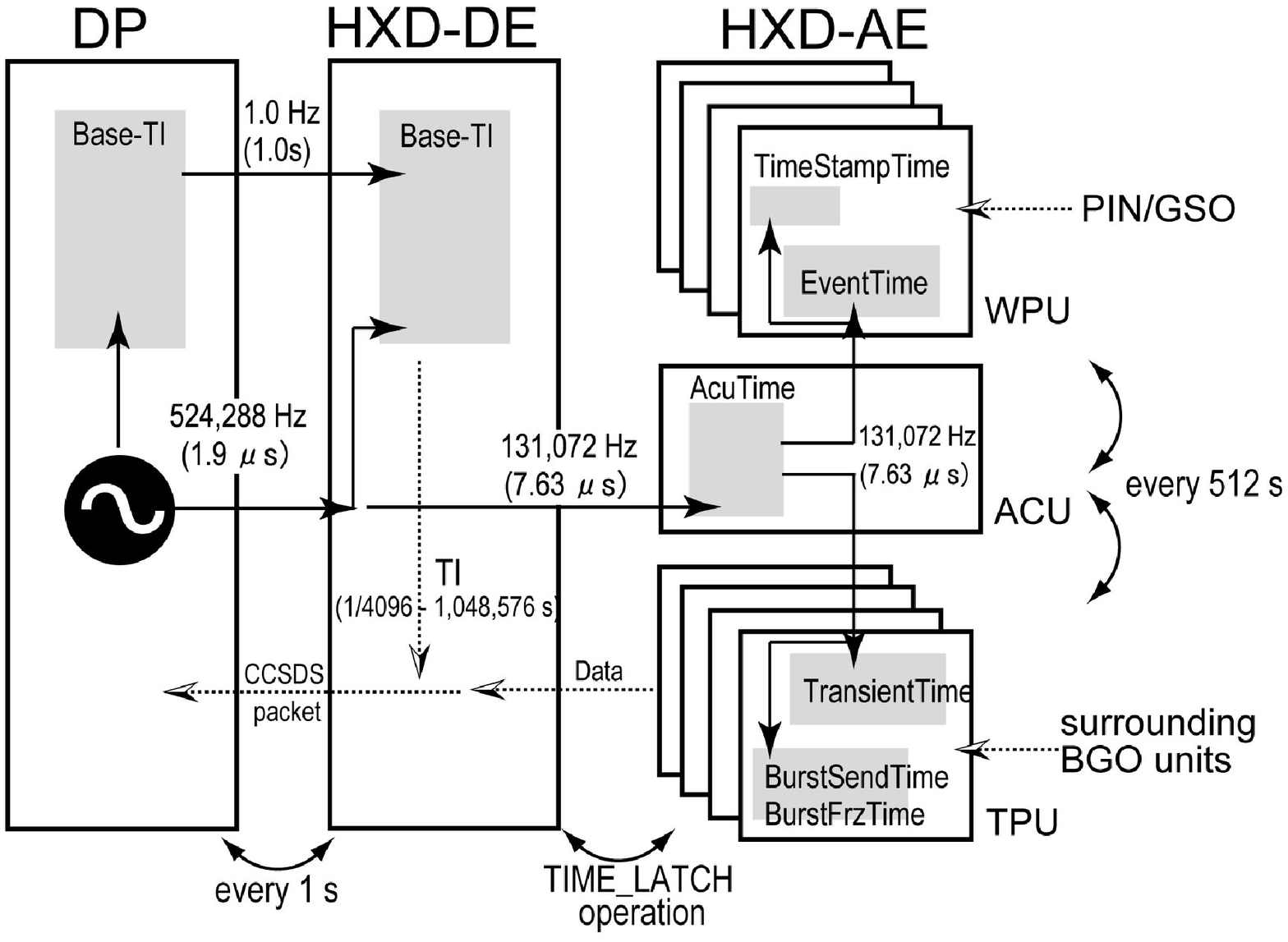}
\end{center}
\caption{Timing counters in the HXD electronics.}
\label{fig:hxd_timing_system}
\end{figure}

Before launch, we repeatedly tested and confirmed the HXD timing capability,
first at each component level, then incorporating HXD-S, HXD-AE, and HXD-DE
\citep{IEEE_kokubun2003,IEEE_tashiro2003,SPIE_kawaharada2004,IEEE_terada2006,IEEE_ohno2006}.
The tests included confirming time assignments of single WEL-type events
with specified time intervals (1, 10, 100, and 1000 s) from a time origin,
detection of periodic signals (with periods of 1, 10, and 100 ms),
and measuring time-interval distribution using random events.
After the HXD was mounted on the spacecraft,
another series of end-to-end tests were repeated incorporating
spacecraft central clock, radio transmitters, and ground receivers.

Through the pre-launch verification,
we confirmed that the HXD has correct timing function as designed.
However, these tests were limited in many aspects.
For example,
the absolute timing measurement was performed
only before  the HXD was mounted on the spacecraft.
The ground equipment used in these end-to-end tests was
not identical to those actually used at the Suzaku tracking center,
the Uchinoura Space Center (USC) in southern Japan.
Furthermore, once the spacecraft is put into orbit,
we must consider additional complications, such as 
signal delays due to air propagation to the ground station,
and due to cable transmission from the antenna to signal receivers.

\begin{figure}[bht]
\begin{center}
\FigureFile(0.95\columnwidth,){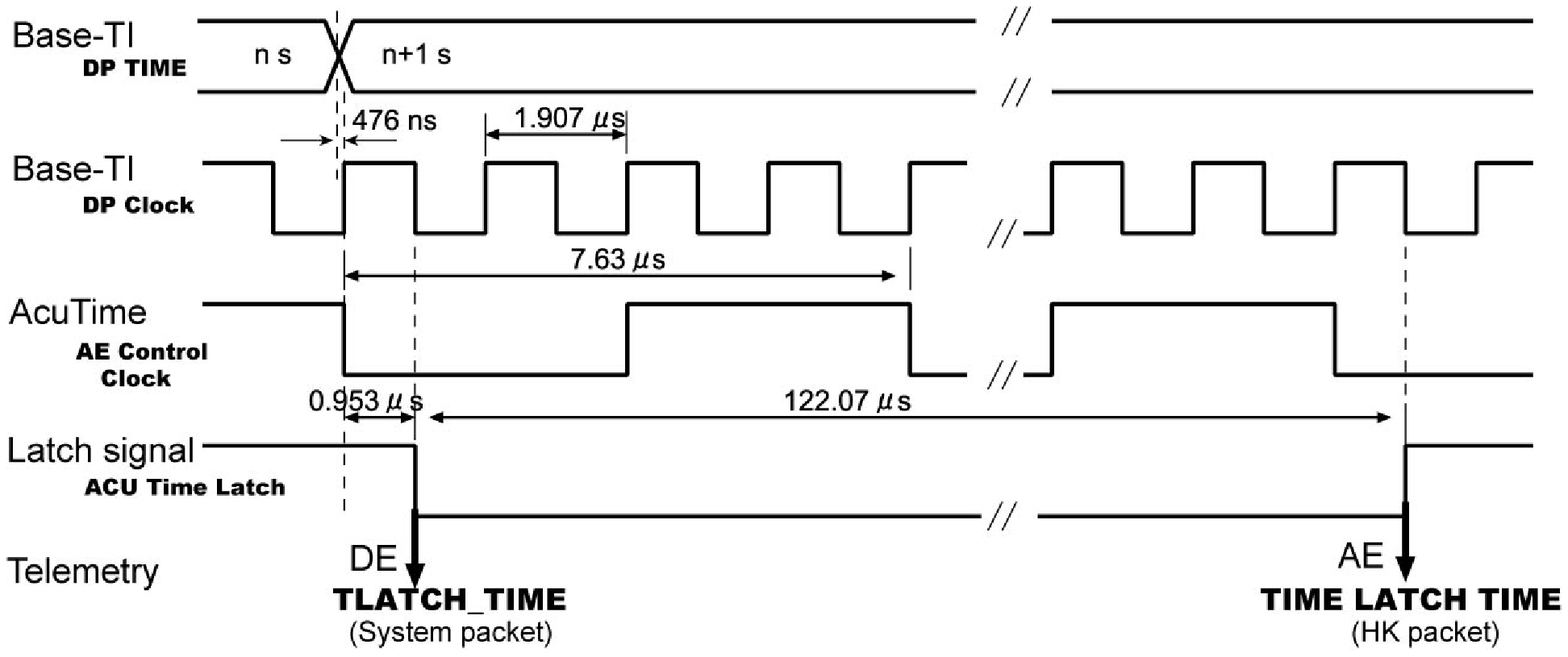}
\end{center}
\caption{A timing chart between HXD-AE and HXD-DE at the TIME LATCH operation. 
DP\_TIME and DP\_Clock are the signals from DP to HXD-DE
to acknowledge time information of the spacecraft,
and AE\_Control\_Clock is provided into HXD-AE from HXD-DE.
The signal named ACU\_Time\_Latch is produced at a time-latch operation
in HXD-DE, and is supplied into HXD-AE to verify timings between them.}
\label{fig:hxd_time_latch}
\end{figure}

The present paper deals with in-orbit timing calibration of the HXD.
After  systematic errors affecting the time assignment
are estimated  in section \ref{section:errors},
we analyze  in section \ref{section:results} highly stable
periodic signals from four fast-rotating pulsars,
including the Crab pulsar in particular.
These tell us the relative accuracy
and stability of the time measurements with the HXD.
Finally, we verify the  absolute timing accuracy of the HXD,
by comparing the HXD measurements of the Crab pulsar
against those with other X-ray missions and with radio telescopes.

\begin{table*}[htb]
\caption{Summary of Timing Counters of the Hard X-ray Detector.}
\label{tbl:hxd_counter}
\begin{center}
\tabcolsep 2pt
\begin{tabular}{llcccl}
\hline 
{Name}&{Component}&{Length}   &{Base}  &{Coverage in Telem.}&{Telemetry Type/Packet}\\
{}    &{}         &{(bit)}    &{(s)}      &{(s)}        &\multicolumn{1}{c}{(HK/Event)}\\
\hline 
{Base-TI} &{DP / Spacecraft} &{39}    &{$1/2^{19}$(1.9$\mu$)}&{$1/2^{12}$(244$\mu$) -- $2^{20}$(1,048,576)}&{HK /time packet}\\
{}        &{HXD-DE}          &{39}    &{$1/2^{19}$(1.9$\mu$)}&{$1/2^{12}$(244$\mu$) -- $2^{20}$(1,048,576)}&{HK / HXD-HK}\\
\hline 
{AcuTime}&{HXD-AE ACU}&{24} &{$1/2^{17}$(7.63$\mu$)}&{$1/2^{15}$(30.5$\mu$) -- $2^{9}$(512) }&{HK / HXD-HK}\\
\hline 
{TimeStampTime}&{HXD-AE WPU}&{24} &{$1/2^{15}$(30.5$\mu$)}&{$1/2^{15}$(30.5$\mu$) -- $2^{9}$(512)}&{HK / HXD-HK}\\
{EventTime$^\dagger$}&{HXD-AE WPU}&{19} &{$1/2^{14}$(61$\mu$)}&{$1/2^{14}$(61$\mu$) -- $2^{5}$(32)}&{Event / HXD\_WEL}\\
\hline 
{TransientTime$^\ddagger$}&{HXD-AE TPU}&{24} &{$1/2^{15}$(30.5$\mu$)}&{$1/2^{15}$(30.5$\mu$) -- $2^{9}$(512)}&{Event / HXD\_WAM}\\
{BurstSendTime}&{HXD-AE TPU}&{23} &{$1/2^{6}$(15.6m)}&{$1/2^{6}$(15.6m) -- $2^{18}$(262144)} &{Event / HXD\_WAM}\\
{BurstFrzTime} &{HXD-AE TPU}&{32} &{$1/2^{15}$(30.5$\mu$)}&{$1/2^{5}$(30.5m) -- $2^{18}$(262144)} &{Event / HXD\_WAM}\\
\hline 
\end{tabular}

\begin{minipage}{1.9\columnwidth}
{\scriptsize
$\dagger$ Time value tagged to each event which detected by a main unit
of the HXD sensor.\\
$\ddagger$ Time value used for the HXD-WAM (Wide-band All-sky Monitor \citep{IEEE_yamaoka2006}).
}
\end{minipage}
\end{center}
\end{table*}

\section{The Timing System of the HXD}
\label{section:hxd_timing}
\subsection{Time Assignment of the Suzaku Data}
\label{section:hxd_timing:suzaku}
All the Suzaku data acquired in orbit utilize a common timing system
based on a 524,288 Hz quartz clock in the spacecraft 
central data processor (DP).
Since the satellite has an orbit with an altitude of 568 km,
an eccentricity of $<0.0002$, and an inclination of 31\fdg4,
it has an orbital period of about 90 min, making 15 revolutions a day.
Therefore, only 5 contacts from the ground station, USC, 
are available per day.
All the data packets are stored into an on-board data recorder (DR) of 6 Gbits,
and transferred via down link of 4.19 Mbps to ground 
during the satellite contacts.
Therefore, the time assignment of the Suzaku data consists of two steps;
the first is to calibrate time origin of the on-board clock 
against the Universal Time Coordinated (UTC) values 
during every ground contact, 
and the second is to determine the time when each telemetry packet was edited
(usually outside the ground contacts), referring to the on-board clock counts
after correction for their temperature-dependent drifts.

\begin{figure}[bht]
\begin{center}
\FigureFile(0.95\columnwidth,){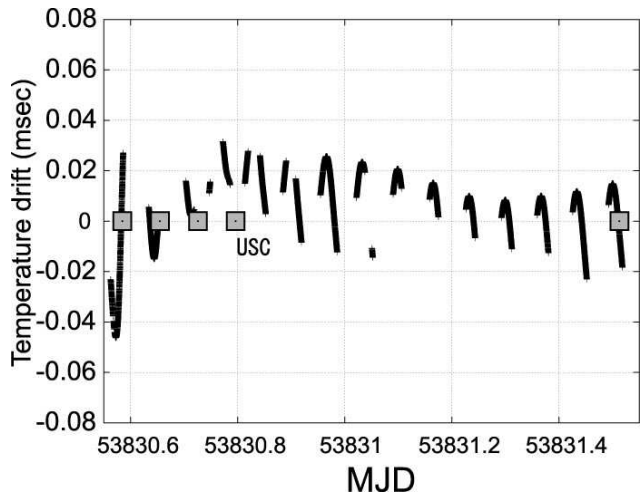}
\end{center}
\caption{An example of cumulative clock drifts, 
expected for the data taken on 2005 September.
They are estimated by the measured DP temperature,
and the temperature-dependent clock drift rate measured before launch.
The horizontal axis shows the Modified Julian Day (MJD).
Satellite contacts at the USC are also indicated.}
\label{fig:cor_temperature}
\end{figure}

In order to afford the first step of the time assignment,
the on-board clock counts, called ``TI'',
are attached to all the data packets (to be sent to telemetry) which are 
in so-called CCSDS (Consultative Committee for Space Data System) format.
Each TI has 32 bits, and covers $10^{20}$ s = 1,048,576 s $\simeq 12$ days 
with 1/4096 s = 244 $\mu$s least significant bit. 
When the spacecraft makes ground contacts, 
this 32-bit values in TI format are calibrated with 
ground Rubidium clocks in UTC format
with a stability of $10^{-11}$s s$^{-1}$ month$^{-1}$.
That is, all the relevant information, 
including the TIs and temperature of the clock,
are gathered every 32 s into special packets called ``timing CCSDS packets'',
which are sent via S-Band real-time down link with the highest priority.
A time delay inside the spacecraft from a packet edition in the DP
to its acknowledge by the spacecraft transmitter is fixed at 18.0 $\mu$s. 
Since the traveling time from the spacecraft to the ground station depends 
on the distance between them, the spacecraft location is measured 
by ranging about twice a day, 
within an accuracy of $1$ km along the line-of-sight, 
while about 10 km in three dimensions.
Finally, a time delay within the ground station, 
from the receiver to the recorder, is also known to be less than 2 $\mu$s.
On each data packet received on ground, a ``time stamps'' is imprinted
based on the Rubidium clocks with a 10 or 100 $\mu$s timing resolution,
depending on which of the 34m and 20m antenna systems at USC, respectively,
is used to receive the data.

In the second step of the time assignment,
TIs issued while the spacecraft is out of contacts are converted to UTC values
through the TI vs.\ UTC relation, established in the first step
using the closest ground contacts.
Since the clock of Suzaku is not placed in a thermostatic environment,
its temperature changes with the spacecraft attitude 
by about $\pm$ 20 K around $\sim 290$ K, 
leading to temperature-dependent clock drifts.
However, this effect has been measured on ground and tabulated 
before the launch, and the table is loaded to the on-board DP.
The DP monitors the clock temperature, and 
calculates every 31.25 ms the corresponding integrated drift time, $Y$, 
in units of 1/4096/4096/100 s = 0.6 ns.
The values of $Y$ are also recorded into the timing CCSDS packets.
According to the pre-launch sea-level tests,
TI values after the temperature correction have a stability of 
about $4\times10^{-9}$, which accumulates to about 25 $\mu$s in one revolution.

\subsection{Calculation of Arrival Time of HXD Events}
\label{section:hxd_timing:hxd}
Using the method described in section \ref{section:hxd_timing:suzaku},
we can determine the time values when individual telemetry packets are edited,
but which we need to know is the arrival time of each event
to the HXD sensor.
Considering a mission requirement of a few hundred $\mu$s resolution,
we designed the HXD so that each event is tagged with its arrival time
information with a resolution of $1/2^{14} = 61.0$ $\mu$s nominally, 
and $1/2^{15} = 30.5$ $\mu$s on condition (selectable by commands).
However, because of severe limitation of the telemetry transfer rate,
only 16 bytes information is available per event,
which must include not only timing, 
but also 6 pulse-heights (4 PINs, GSO, and BGO), quality flags, 
trigger information, and hit-pattern information 
\citep{SPIE_takahashi1998,IEEE_terada2006}.
Consequently, only 19 bits per event (hereafter ``EventTime'') 
are available for the timing, 
which covers $1/2^{14} = 61.0$ $\mu$ to 32 seconds.
Thus, we need to have a mechanism to relate ``EventTime''
to the spacecraft time in TI.

Figure \ref{fig:hxd_timing_system} shows various timing counters 
in the HXD electronics, which are also summarized 
in table \ref{tbl:hxd_counter}.
The spacecraft clock (section \ref{section:hxd_timing:suzaku})
stays in the data processing unit (DP), and generates a timing clock 
of $1/2^{19} = 1.9$ $\mu$s period. 
As shown there, the clock is first supplied to HXD-DE, then 
divided into the control module (Analog Control Unit, ACU) of HXD-AE, 
and finally supplied into eight analog modules 
(four Well-type detector Processing Unit, WPU, 
and four Transient detector Processing Unit, TPU) of HXD-AE.

HXD-DE always holds the latest TI value, supplied from DP 
every $2^{19}$ DP clocks (corresponding to 1 s).
In HXD-AE, the time reference is provided by ``AcuTime'' in ACU,
which is refreshed every 512 s by carry reset signals. 
The values of counters of ``AcuTime'' and TI are synchronized, and 
cross-checked at the beginning of each observation via TIME\_LATCH operation,
where they are latched according to the timing chart in 
figure \ref{fig:hxd_time_latch}. The latched values of ``AcuTime'' and TI 
are sent to the telemetry, and stored as 
`HXD\_AE\_TM\_LATCH\_TM' and `HXD\_TLATCH\_TIME' columns 
in the HXD HK FITS file, respectively.
This operation occurs typically once every one to two days.
Besides, the eight analog precessing modules in HXD-AE 
(four WPUs and four TPUs) 
have such timing counters as ``TimeStampTime'', ``EventTime'', 
``TransientTime'', ``BurstSendTime'', and ``BurstFrzTime'',
which are synchronized to the ``AcuTime'' signals.
 
As explained so far, 
an X-ray event detected with the HXD is given a UTC time through the following
two steps;
first to relate a value of 19-bit ``EventTime'' into 
the corresponding TI value, and then to convert it into the UTC value 
as described in section \ref{section:hxd_timing:suzaku}.
In order to conduct the first step 
without any ambiguity due to scalar overflows, 
the timing system of the HXD is designed to ensure that 
the processing and waiting time from the detection of an event to 
its telemetry output is less than 32 s, which is the coverage of ``EventTime''.
The algorithm of the HXD time assignment is implemented in the 
analysis tool, ``hxdtime'', in the HEAsoft package.

\section{Estimation of errors in the time assignment}
\label{section:errors}
\subsection{Effects of Spacecraft Location Errors}
\label{section:errors:usc_orbit}
As described in section \ref{section:hxd_timing:suzaku},
the first step of the time assignment of Suzaku 
is affected by systematic errors 
in a determination of orbital parameters.
Due to these errors, 
the instantaneous spacecraft location has uncertainties of about 10 km,
whereas errors along the line-of-sight from the ground station 
to the satellite are much smaller, about 1 km. 
Therefore, the air propagation delay of the telemetry 
can be calculated to an accuracy of 3 $\mu$s or so.

\begin{figure}[bht]
\begin{center}
\FigureFile(0.9\columnwidth,){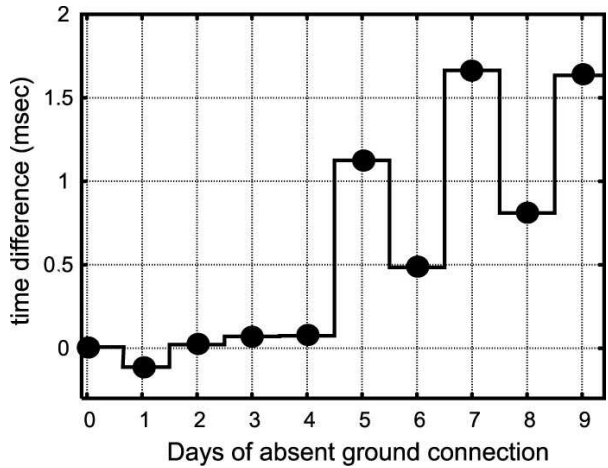}
\end{center}
\caption{Time assignment errors introduced to the Crab Nebula data
taken on 2003 August 26, when the TI to UTC calibration 
in the off-line data processing is purposely discarded 
for various length of time.
The horizontal axis shows duration without ground contacts.
The attitude of the satellite was changed on the forth day 
in the plot to observe another object.}
\label{fig:chg_pass_summary}
\end{figure}

In astrophysical analyses of the HXD data, 
the orbital parameters of Suzaku are also used for another purpose;
to perform so-called barycentric correction, or 
to convert the arrival times of photons from celestial bodies
for light-travel times between Suzaku and 
the center of gravity of the Solar system.
We have developed such a tool, named {\it aebarycen} 
in HEAsoft package, to perform the barycentric correction on each event.
The correction has three steps; 
first, using the orbital parameters of the spacecraft, 
event arrival times are converted to those to be measured
at the geodetic center of the Earth, 
second, to the center of the Sun, 
and finally to the center of the Solar system.
The largest conversion in these steps comes from the second one,
which is an order of a few hundred s (maximum 8 min
corresponding to the traveling time from the Sun to the Earth), 
although it can be calculated accurately from the position of the object
and the observation date (i.e, the position of Sun and the earth).
Systematic errors in the overall barycentric correction mainly arise 
in the first step;
the errors of $\sim$10 km associated with the satellite position
causes the barycentric times to be uncertain by $\sim 30$ $\mu$s.

\begin{figure}[bht]
\FigureFile(0.9\columnwidth,){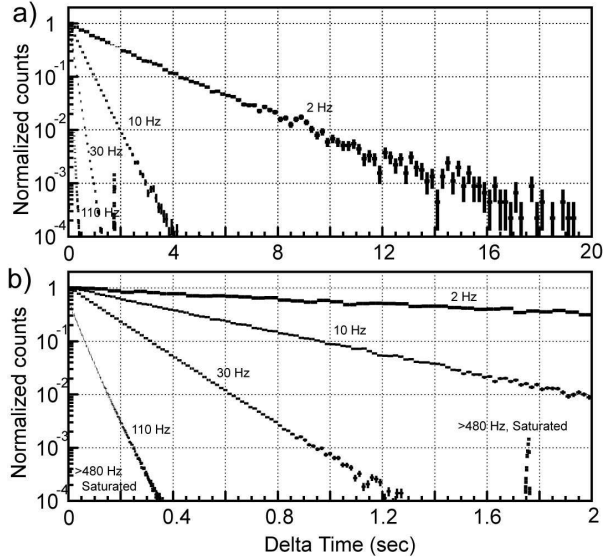}
\caption{Examples of time-interval histograms of HXD WEL events
at various counting rates, 2, 10, 30, 110, and $>$480 Hz.
The 2 Hz dataset was derived from cleaned PIN events 
acquired on 2005 September 14 for 21 ks, and 
those of 10 and 30 Hz are from GSO events of 
Crab observation on 2007 March 20
when the telemetry mode was set for bright sources.
Only a quarter of GSO events were sent to telemetry
when the 10 Hz data were acquired.
The 110 Hz data set was
taken on 2006 May 17 for an exposure of 76 ks
(observation of Lockman Hole), and those with $>$480 Hz was taken 
on 2005 September 14
during a daily health-check operation of the PSD function 
\citep{SPIE_takahashi1998,SPIE_tanihata1999}
in which the on-board background rejection was temporally disabled.}
\label{fig:study_delta_time}
\end{figure}

\subsection{Effects of Temperature Variation on the Clock}
\label{section:errors:temperature}
At the second step of the time assignment of data packets 
(section \ref{section:hxd_timing:suzaku}), 
namely in the conversion of TIs to UTCs,
the clock frequency must be corrected for temperature-dependent drifts.
Although the clock stability after the correction 
was confirmed to be $4\times10^{-9}$ 
in a pre-flight test (section \ref{section:hxd_intro}),  
the relation between the drift rate and DP temperature,
used in this correction, was calibrated only on ground,
and needs an in-orbit reconfirmation.
Before detailed studies in section \ref{section:results},
figure \ref{fig:cor_temperature} shows an example of 
expected clock drifts to be accumulated in $\sim$ 100 minutes;
the drift is thus predicted to be typically about 20--40 $\mu$s,
which introduces systematic errors in the absolute timing 
by the same order of magnitude in the worst case.

\subsection{Effects of Intervals between Ground Contacts}
\label{section:errors:usc}
The cross-comparison of TIs to UTC values are performed 
only at ground contacts.
Consequently, systematic errors associated with the time-assignment on packets 
depend mainly on the time intervals between adjacent contacts.
The operations at the down-link station are sometimes canceled 
due, e.g., to launch campaigns of new satellites,
or typhoons attacking the local area.
To emulate the effect of loss of ground contacts, we intentionally, 
in an off-line data analysis, omitted the TI vs. UTC cross calibrations
for a given length of days.
Then, as shown in figure \ref{fig:chg_pass_summary}, the cumulative timing
errors increased as the assumed duration without ground contacts get longer.

\section{Results of In-Orbit Timing Calibrations}
\label{section:results}
\subsection{Time Assignment of Random Signals}
\label{section:results:random}
As described in section \ref{section:hxd_timing:hxd},
the data acquisition system of the HXD is designed to avoid
internal data buffering for longer than 32 s.
Any failure in the design would produce artifacts in 
``time-interval'' spectra (occurrence distributions of time difference
between adjacent events), which would normally be a simple exponential 
when the signals are random.
We selected in-orbit unfiltered-WEL events (i.e., which include 
all of PIN, GSO, and BGO events)
at various counting rates, and produced time-interval histograms. 
As shown in figure \ref{fig:study_delta_time},
the derived spectra all exhibit straight lines 
in semi-log plots, except when the telemetry is saturated.
When the HXD event rate becomes so high 
that the required data flow rate exceeds the telemetry limit,
some packets are discarded on HXD-DE every fixed time
(1, 2, 4, or 16 s when the data rate of the telemetry is
Super High, High, Medium, or Low), and thus, 
an artificial structure appears in the time-interval spectra;
this is seen at 1.7 s in figure \ref{fig:study_delta_time}
for the 480 Hz case.

\begin{table*}[hbt]
\caption{Pulsars observed with Suzaku.}
\label{tbl:suzaku_pulsars}
\begin{center}
\begin{tabular}{llcc}
\hline 
{Name}&{Period}&{Observation date$^\dagger$}&{Net Exposure of the HXD}\\
\hline 
{Crab}&{33.58ms}&{2005/08--2007/03}      &{200ks (47 observations)}\\
{PSR1509-58b}&{151.3533 ms}&{2005/08/23}&{44.6 ks}\\
{Hercules X-1}&{1.237 s$^\ddagger$}&{2005/10/05,2006/03/29}&{30.7 and 34.4 ks}\\
{A0535+262}&{103.375 s}&{2005/09/14}&{21.3 ks}\\
\hline 
\end{tabular}
\begin{minipage}{1.9\columnwidth}
{\scriptsize
$\dagger$ In the format of year/month/day;\hspace{10pt} 
$\ddagger$ Without correcting for the Doppler shifts due to 
binary motion of the object.}
\end{minipage}
\end{center}
\end{table*}

\subsection{Timing Verification with Periodic Signals}
\label{section:results:periodic}
\begin{figure*}[htb]
\begin{center}
\FigureFile(1.7\columnwidth,){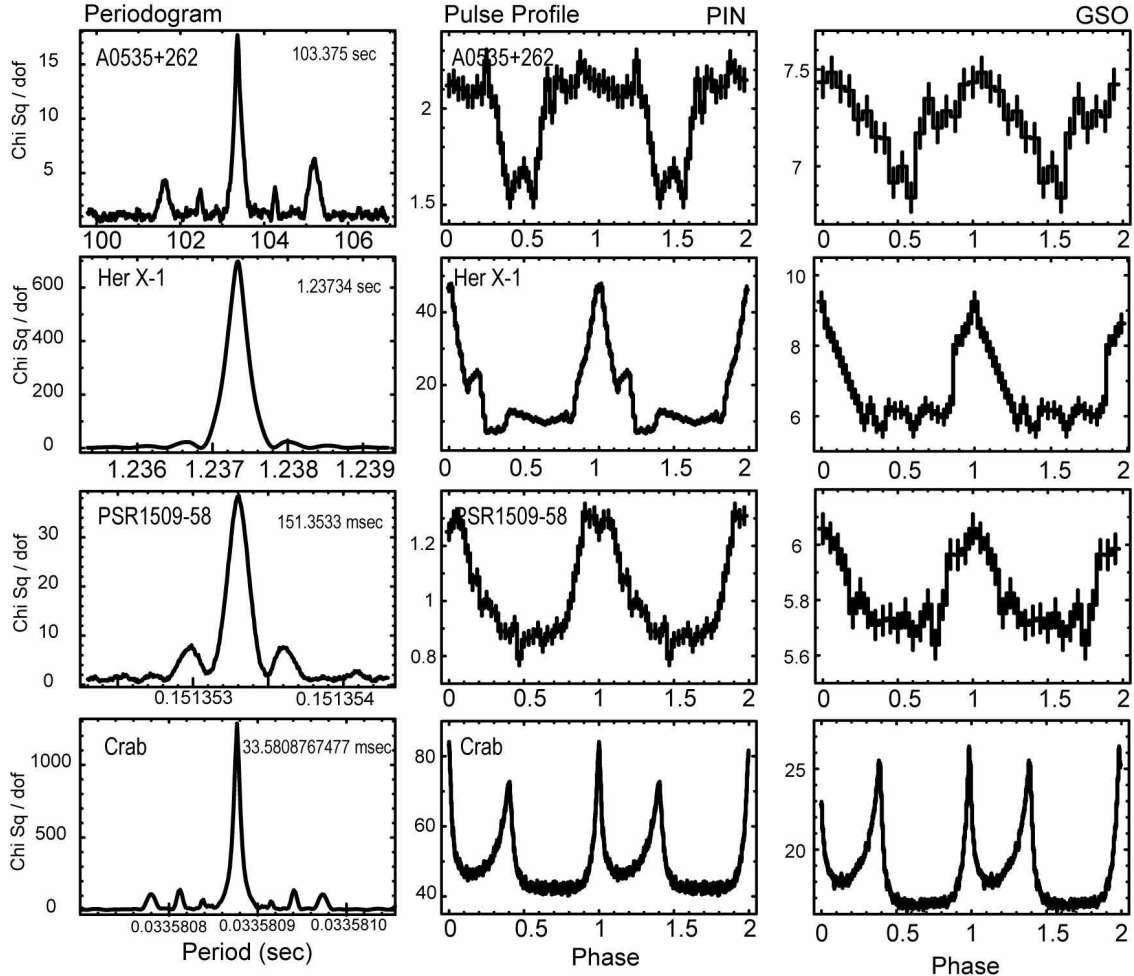}
\end{center}
\caption{Periodograms and pulse profiles of A0535+262, Hercules X-1, 
PSR1509$-$58, and the Crab pulsar, observed with the HXD. 
Left panels show the periodograms, obtained by PIN in the 10--70 keV band. 
The vertical axis means a chi-squared value when a light curve, 
folded at trial periods shown in the horizontal axis, 
is fitted by a constant. Middle and right panels show pulse profiles
by PIN and GSO, respectively, folded at the periods determined by PIN 
in the left panel.}
\label{fig:pulsars_pin}
\end{figure*}

\begin{figure}[htb]
\begin{center}
\FigureFile(0.90\columnwidth,){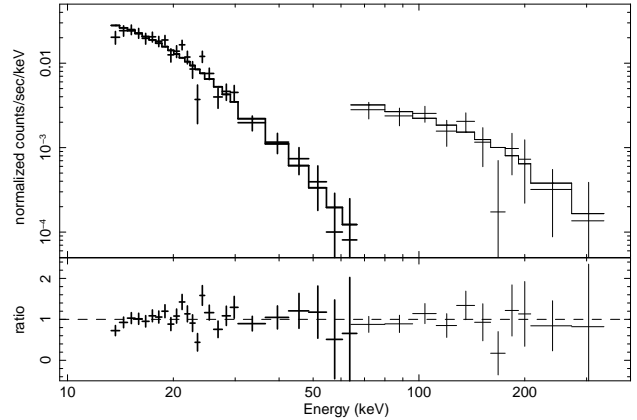}
\end{center}
\caption{Pulsed-component spectra of PSR1509-58, obtained with the HXD
by subtracting the pulse-bottom spectra from the one at the pulse peak,
reffering to the pulse profile shown in figure \ref{fig:pulsars_pin}.
The pulse peak and bottom is defined as a pulse phase of 
0.0 $\pm$ 0.2 and 0.5 $\pm$ 0.3, respectively.
The data of PIN and GSO are shown in croses, and 
best fit power-law model of them are shown with solid lines.
The fitting ranges are 13--70 keV and 70--300 keV band 
for the PIN and GSO data, respectively.}
\label{fig:psr1509_pulse}
\end{figure}

In the initial performance verification phase of Suzaku,
we observed four pulsars listed in table \ref{tbl:suzaku_pulsars}.
After the barycentric corrections (section \ref{section:errors:usc_orbit}), 
we have successfully detected the periodic signals from them,
as shown in periodograms of figure \ref{fig:pulsars_pin}.
As already reported for several of them \citep{A0535_apjl,herx1_pasj},
pulse profiles of these objects observed with the HXD are consistent
with those measured in previous observations.
The pulsations were also detected in the GSO data of these objects.
As shown in figure \ref{fig:psr1509_pulse}, 
energy spectra of the pulse component of PSR1509-58 can be reproduced 
by a single power-law model up to 300 keV or higher.
The best-fit parameters are a photon index of $1.55_{-0.09}^{+0.10}$ and
X-ray flux of $4.86_{-0.48}^{+0.34} \times 10^{-10}$ erg s$^{-1}$ cm$^{-2}$
in the 10 -- 300 keV band.
Thus, the HXD has a capability of detecting periodic signals 
in the energy band of 10 to a few hundred keV, 
from an object of 100 mCrab intensity only in a 45 ks exposure.

\begin{figure}[bht]
\begin{center}
\FigureFile(0.9\columnwidth,){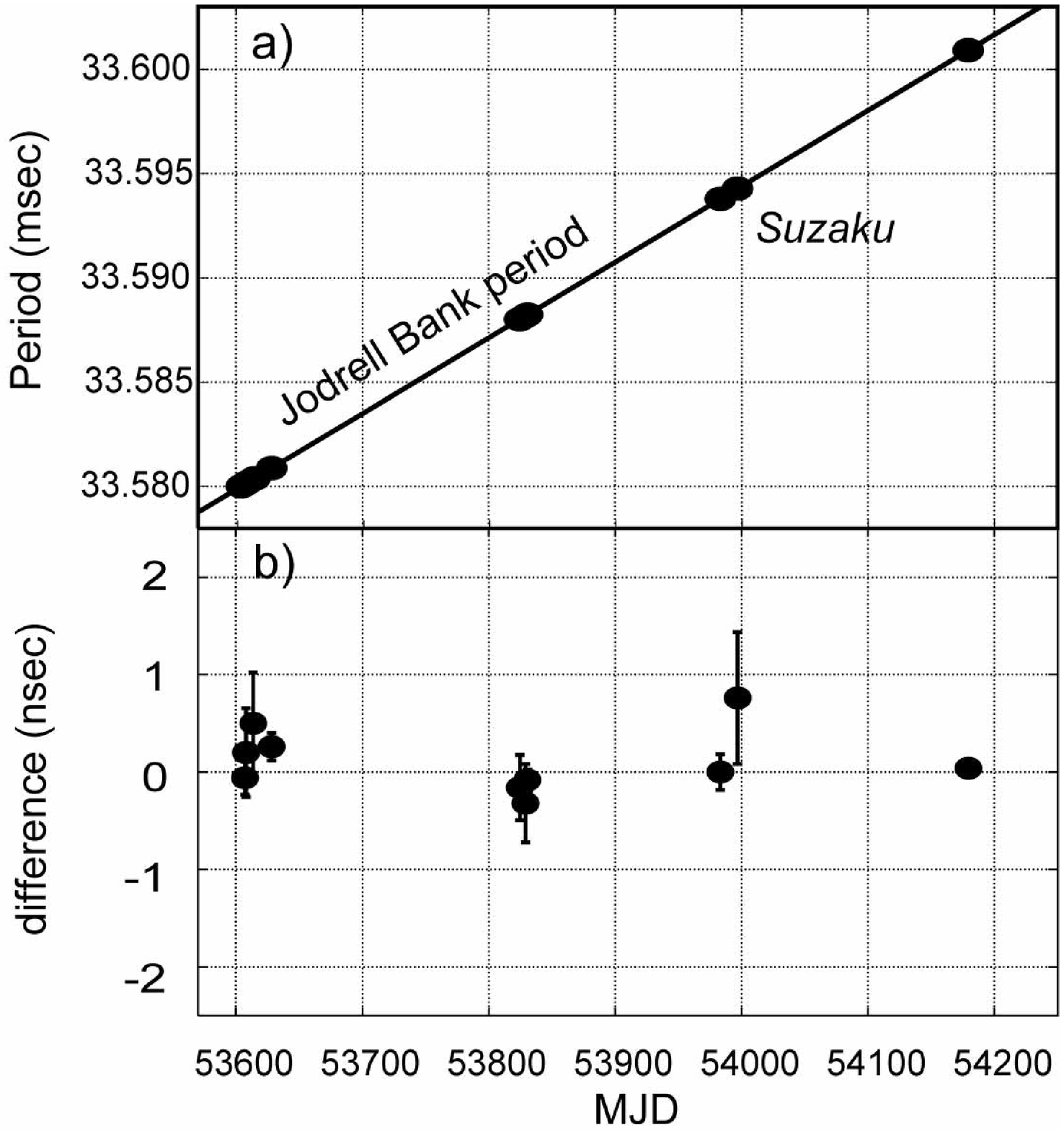}
\end{center}
\caption{Pulsation periods of the Crab pulsar obtained by the HXD, 
compared with the radio measurements (the solid line) 
at the Jodrell Bank observatory \citep{JODRELL_RADIO_CRAB}.
The residuals are shown in the bottom panel.
Error bars were determined by the method by \citet{period_err} 
with a single iteration, and refer to 1 sigma.}
\label{fig:crab_obs_period}
\end{figure}

\subsection{Relative and Absolute Timing Calibration with Crab}
\label{section:crab}
As listed in table \ref{tbl:suzaku_pulsars}, 
we observed the Crab pulsar 47 times for several calibration purposes
including timing, absolute flux, energy spectral shape, and angular response.
Among the 47 observations, we selected 23 observations, which have 
exposure longer than 20 ks, and 
good statistics to show high power in the period search.
As plotted in figure \ref{fig:crab_obs_period} (top), 
the pulse periods measured on these 23 observations reveal the well know
spin-down trend of the Crab pulsar, which is determined to be 
$4.16 \pm 0.02 \times10^{-13}$ s s$^{-1}$ with our data.
This rate agrees with the results by continuous radio-monitoring 
observations \citep{JODRELL_RADIO_CRAB}.
In addition, as shown in figure \ref{fig:crab_obs_period} bottom,
the individual periods obtained by the PIN datasets 
agree with those measured by the Jodrell Bank Radio observatory 
\citep{JODRELL_RADIO_CRAB} within 1.0 ns. 
The error bars shown here, at 1 $\sigma$, were determined in 
reference to \citet{period_err} considering higher harmonics.

\begin{figure*}[hbt]
\begin{center}
\FigureFile(1.8\columnwidth,){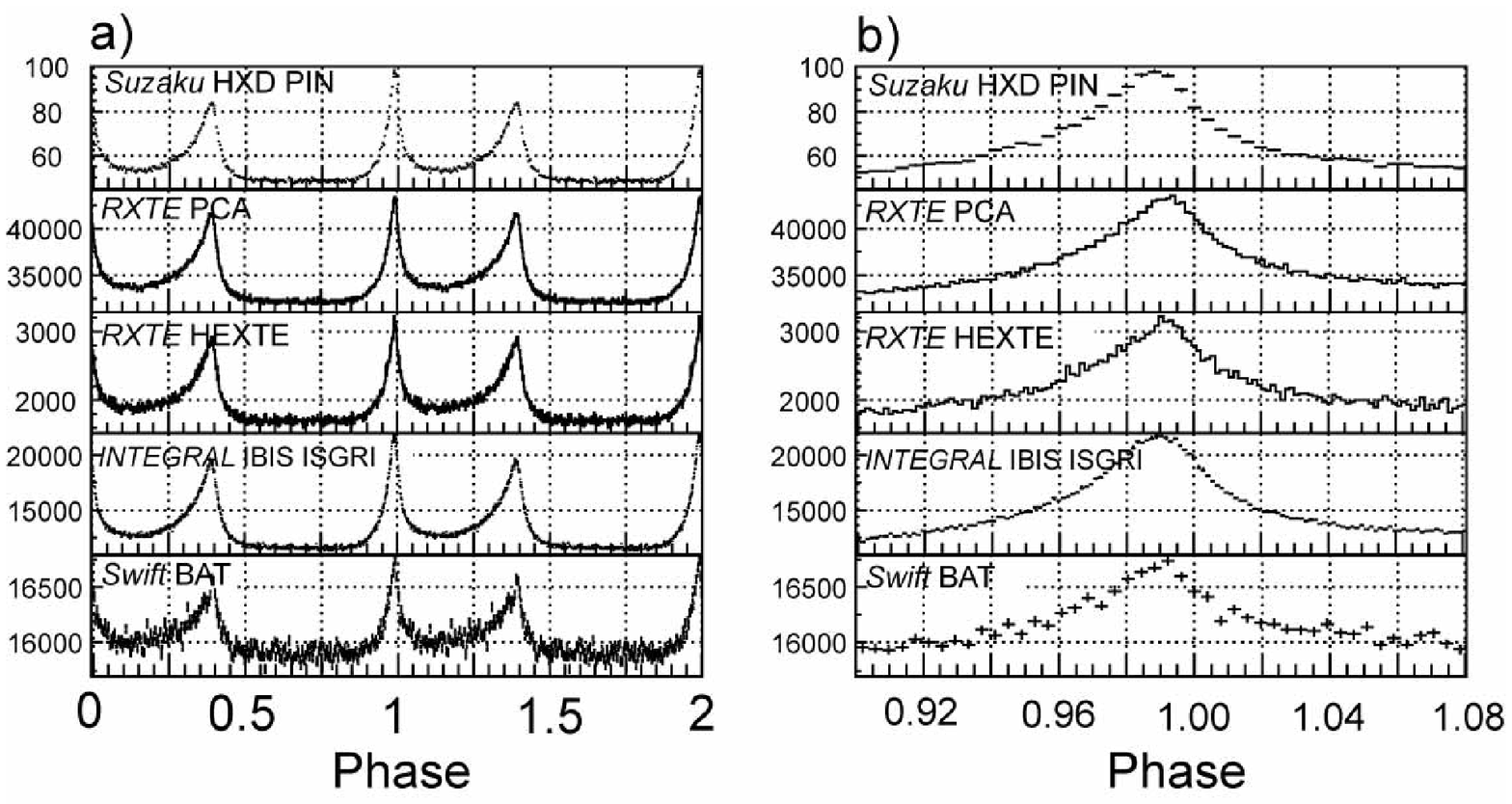}
\end{center}
\caption{X-ray pulse profiles of the Crab pulsar obtained by the Suzaku HXD PIN (10--70 keV), the RXTE PCA (2--30 keV), the HEXTE (20--100 keV), the INTEGRAL IBIS ISGRI (20--100 keV), and the Swift BAT (15--350 keV) from top to bottom. The right panel shows the same plot but expanded to reveal details over the phases of 0.91 to 1.08. The period is 0.03360091293 s, $\dot{P}$ is 4.20506179 $\times 10^{-13}$ s s$^{-1}$, and the phase 0 corresponds to the arrival time of the first pulse obtained by the Jodrell Bank Radio observatory; MJD 54179.4400000438310185391 (where MJD in Terrestrial Time).}
\label{fig:crab_simultaneous_pulse}
\end{figure*}

\begin{figure*}[hbt]
\begin{center}
\FigureFile(1.8\columnwidth,){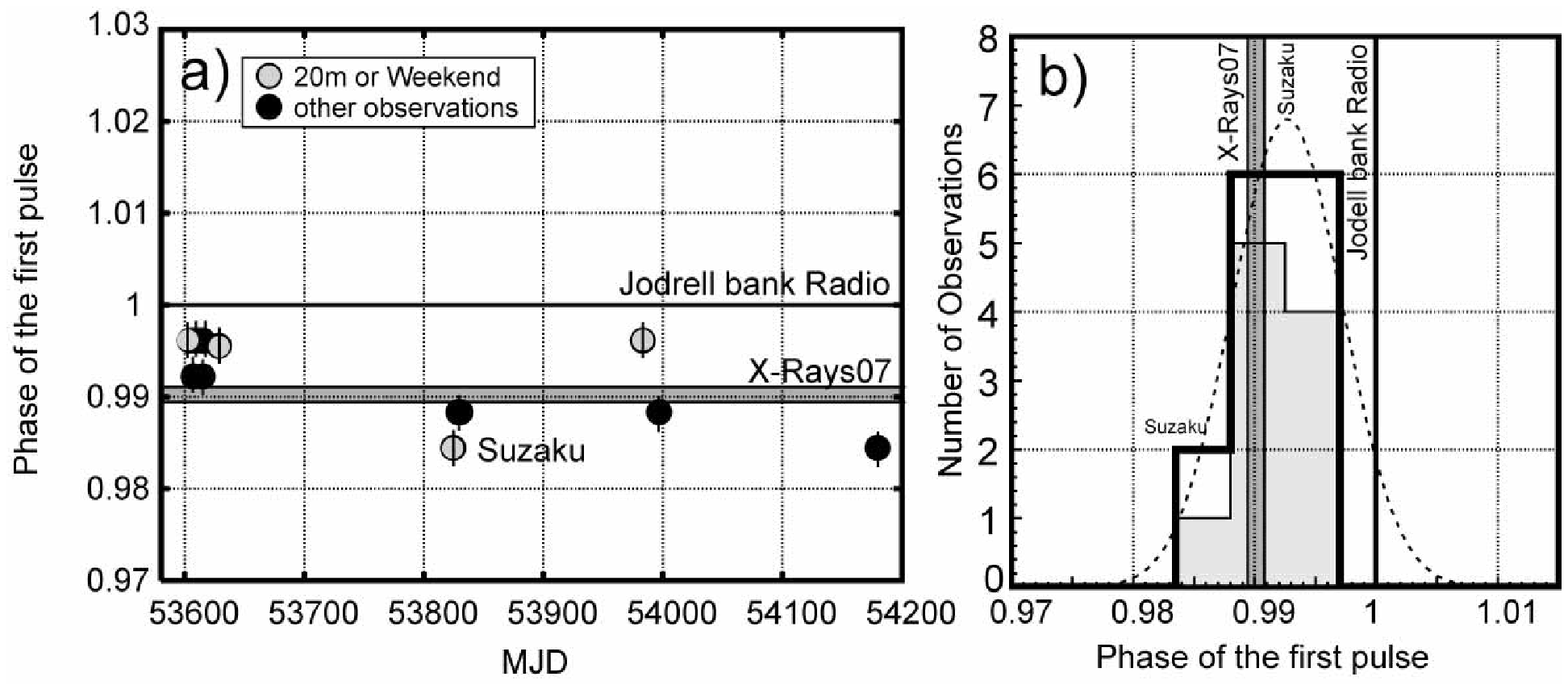}
\end{center}
\caption{(Left)Phases of the first pulse of Crab obtained by HXD PIN, compared to the Jodrell Bank ephemeris \citep{JODRELL_RADIO_CRAB}. Phase 0.0 corresponds to the barycentric arrival time of the radio pulses determined by the Jodrell Bank observatory, and the Suzaku data are shown in circles. Open-gray circles show results when the observation was done with the 20m antenna or across a lack of ground contacts for $>$ 1 day. Filled circles was taken with the 34m system on working days. The phases tagged as ``X-ray 07'' are determined in figure \ref{fig:crab_simultaneous_pulse} by the X-ray missions in the simultaneous observation on 2007 May. 
(Right) A distribution (thick histograms) of the 14 phase measurements of the Crab pulse made with the HXD, compared with the radio ephemeris. The shaded histogram shows a distribution of 10 selected observations, which are taken with 34m antena under daily ground contacts. The dashed curve is a Gaussian fitted to the distribution of all 14 phase measurements.}
\label{fig:crab_arrival_phase}
\end{figure*}

The absolute timing of the HXD can be verified by 
comparing arrival times of the main pulse of the Crab 
with those indicated by the Jodrell Bank Radio ephemeris 
\citep{JODRELL_RADIO_CRAB}.
Thus, on 2007 March 20-21, 
we have performed a simultaneous observation of Crab 
with other X-ray missions, {\it RXTE}, {\it INTEGRAL}, and {\it Swift}. 
The net exposure of the HXD was 41.1 ks,
of which the overlaps with {\it RXTE}, {\it INTEGRAL}, 
and {\it Swift} were 14.5, 81.3, and 24.8 ks, respectively.
The pulse profiles of Crab taken in the campaign are summarized in 
figure \ref{fig:crab_simultaneous_pulse}.
Thus, the X-ray pulses arrive systematically earlier by 
$\sim$ 340--500 $\mu$s than those in the radio band.
This reconfirms a previous report by \citet{crab2004},
who measured the same quantity as 344 $\mu$s with RXTE,
and discussed possible astrophysical implications.
From cross-correlation studies among pulse profiles,
the main pulse measured with the Suzaku HXD precedes, 
by $0.003\pm0.001$, $0.002\pm0.001$, $0.002\pm0.001$, $0.002\pm0.001$
phases, those from the RXTE PCA, the HEXTE, the INTEGRAL IBIS ISGRI, and
the Swift BAT, respectively. 
Thus, in this observation, the absolute timing with the HXD 
systematically leads by about 160 $\mu$s those of the other X-ray instruments. 

The time difference between the HXD and 
the other high energy experiments in the Crab simultaneous observation
is consistent with systematic errors in the HXD timing. 
We estimate these systematic errors 
by comparing several HXD-measured arrival phases of the main pulse of Crab, 
with the Jodrell Bank Ephemeris, 
as shown in figure \ref{fig:crab_arrival_phase}.
The averaged phase of 0.993 is consistent with the other X-ray instruments 
on the simultaneous observation.
Since the phases measured with the HXD thus scatters by about 
$0.010\pm0.005$ phases (with 90 \% error), 
the total systematic errors in the absolute timing of the HXD is concluded 
to be about $360\pm150$ $\mu$s.
If we discard datasets taken with the 20m antenna system or 
those obtained across a lack of ground contacts for $>$ 1 day,
the dispersion reduces to $270\pm130$ $\mu$s.
Although the particular case of figure \ref{fig:crab_simultaneous_pulse}
revealed a systematic offset (by $\sim 160$ $\mu$s) of the HXD timing
from those of other instruments, 
this is within the estimated systematic error of $\sim 270$ $\mu$s.
On the other hand, the phases of the main pulses stay within 0.005 phases
within an observation, 
implying the stability of arrival time measurements 
is within $1.9 \times 10^{-9}$.

\section{Conclusion}
Systematic errors on the HXD timing due to spacecraft location errors,
the temperature-dependence of the on-board clock, 
and between ground contacts (if it is less than 5 days), 
are $\sim 30$, $<40$, and $<150$ $\mu$s, respectively.
Using the Crab pulsar, we confirmed
the stability of the relative timing 
to be $1.9 \times 10^{-9}$ s s$^{-1}$ per day. 
The systematic errors on the HXD absolute time have been calibrated 
as about 360 $\mu$s or higher ($\sim$ 270 $\mu$s 
under daily ground contacts with the 30m antenna system).

\section*{Acknowledgements}
They would like to thank all the members of
the Suzaku Science Working Group,
for their contributions in the instrument preparation,
spacecraft operation, software development,
and in-orbit instrumental calibration.
They also deeply thank 
Dr.\ Richard Rothschild, Dr.\ Keith Jahoda, Dr.\ Arnold H. Rots, 
Dr.\ Christoph Winkler, Dr.\ Wim Hermsen, Dr.\ Erik Kuulkers, 
Dr.\ Jean Swank, Dr.\ Goro Sato, Dr.\ Takanori Sakamoto, 
and the operation teams of {\it RXTE}, {\it INTEGRAL}, and {\it Swift},
for their efforts on the simultaneous observation of Crab
and quick analyses of the data.


\clearpage


\end{document}